\documentclass[sigconf]{acmart}
\AtBeginDocument{%
  }

\copyrightyear{2025}
\acmYear{2025}
\setcopyright{rightsretained}
\acmConference[UIST Adjunct '25]{The 38th Annual ACM Symposium on User
Interface Software and Technology}{September 28-October 1, 2025}{Busan,
Republic of Korea}
\acmBooktitle{The 38th Annual ACM Symposium on User Interface Software and
Technology (UIST Adjunct '25), September 28-October 1, 2025, Busan, Republic
of Korea}\acmDOI{10.1145/3746058.3758410}

\acmISBN{979-8-4007-2036-9/2025/09}

\begin{document}

\title{Your Thoughtful Opponent: Embracing Cognitive Conflict With a Peer Agent}


\author{Kyuwon Kim}
\affiliation{%
  \institution{Ewha Womans University}
  \city{Seoul}
  \country{Republic of Korea}}
\email{kyuwonkim95@ewha.ac.kr}

\author{Jaeryeong Hwang}
\affiliation{%
  \institution{Ewha Womans University}
  \city{Seoul}
  \country{Republic of Korea}}
\email{jryeong67@naver.com}

\author{Younseo Lee}
\affiliation{%
  \institution{Ewha Womans University}
  \city{Seoul}
  \country{Republic of Korea}}
\email{03dldbstj0915@gmail.com}

\author{Jeanhee Lee}
\affiliation{%
  \institution{Ewha Womans University}
  \city{Seoul}
  \country{Republic of Korea}}
\email{jinnylee26@ewha.ac.kr}

\author{Sung-Eun Kim}
\affiliation{%
  \institution{Ewha Womans University}
  \city{Seoul}
  \country{Republic of Korea}}
\email{sungeunkim@ewhain.net}

\author{Hyo-Jeong So}
\affiliation{%
  \institution{Ewha Womans University}
  \city{Seoul}
  \country{Republic of Korea}}
\email{hyojeongso@ewha.ac.kr}
\renewcommand{\shortauthors}{Kim et al.}

\begin{abstract}
As complex societal issues continue to emerge, fostering democratic skills like valuing diverse perspectives and collaborative decision-making is increasingly vital in education.  In this paper, we propose a Peer Agent (PA) system designed to simulate a deliberative conversational partner that induces socio-cognitive conflict within dilemma-based game play. Drawing on by the Inner Thoughts framework and grounded in value-sensitive discourse analysis, the PA actively participates in voice-based multi-party deliberation with human players. The system architecture consists of five core modules: Context Interpreter, Agent State Manager, Thought Generator, Thought Evaluator, and Thought Articulator.
\end{abstract}

\begin{teaserfigure}
  \includegraphics[width=\textwidth]{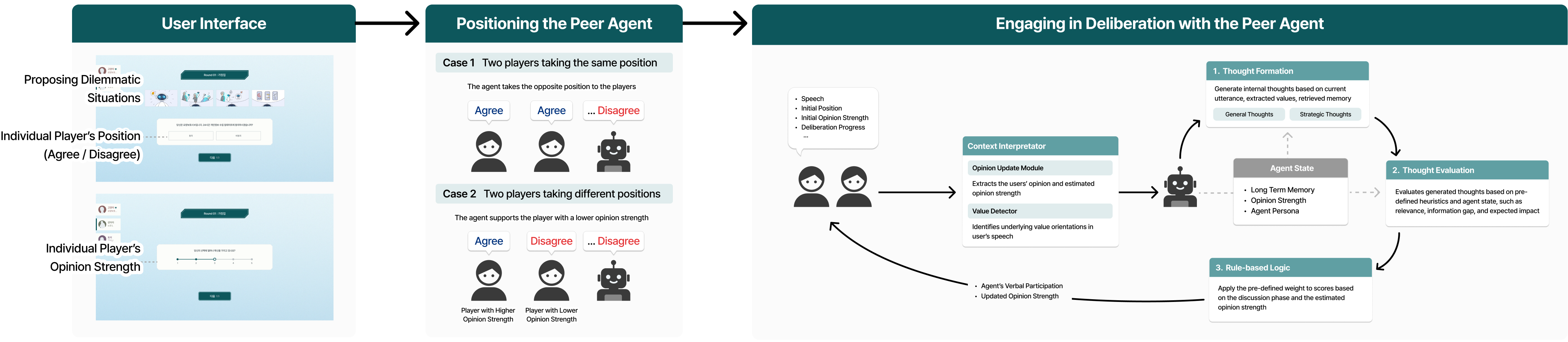}
  \caption{Proposed interface and overall workflow of our peer agent system. This system aims to foster socio-cognitive conflict throughout the deliberation process by introducing diverse viewpoints which learners overlook.}
  \Description{The figure illustrates the proposed interface and overall workflow of the peer agent system. After receiving the user's input, the agent adopts the user's position and begins engaging in the deliberation process.}
  \label{fig:teaser}
\end{teaserfigure}

\maketitle

\section{INTRODUCTION}
As society becomes increasingly complex and diverse, new moral and controversial issues continue to emerge, underscoring the importance of reaching a socially acceptable consensus. However, achieving such consensus remains challenging due to the involvement of various stakeholders with conflicting values and perspectives \cite{rudschies2021value}. Integrating collaborative decision-making practices into technology-enhanced learning environments allows for the simulation of real-world problems within a safe space to support learners to explore complex issues \cite{Elder_1971}. Exposure to differing viewpoints often leads to socio-cognitive conflict, which is a form of cognitive tension that can promote cognitive restructuring and deeper understanding \cite{doise1979individual, mugny1978socio}.

While collaborative decision-making has the potential to support deeper and logical thinking, such deliberative interactions rarely occur in natural settings. To facilitate constructive deliberation, LLM-powered deliberation systems have emerged as a promising area of research \cite{tessler2024ai, lee2025amplifying}. However, most of the existing work has focused on the role of LLMs in synthesizing or amplifying pre-existing opinions, rather than leveraging LLMs to simulate and introduce diverse perspectives \cite{nguyen2024simulating}. In this paper, we propose a peer agent (PA) system that stimulates players' thinking through structured deliberation during dilemma-based game play. PA has the potential to induce productive disagreement in discussions due to its active role-switching features \cite{cassell2022socially}. We present the design of the PA system architecture and a preliminary user study to explore the opportunities and challenges of our proposed approach.

\section{SYSTEM DESIGN}
\subsection{Proposed Interface and Workflow}
Figure 1 outlines the proposed interface and workflow of our system, which simulates a multi-party deliberation involving two human players and one PA with the goal of achieving consensus on AI ethics dilemmas. Our system incorporates a voice user interface since voice is the most natural modality of human communication, and can facilitate natural turn-taking and interaction in group conversations with an agent \cite{reicherts2022s}.

The system interface of our dilemma-based game with three players presents AI ethics-related issues (e.g., should we allow the development of AI killer robots?), forcing players to respond to two queries: (1) their stance on the proposed dilemma (Agree or Disagree), and (2) their subjective confidence in the selected stance rated on a 1-5 scale to indicate the strength of their opinion. These two queries serve as core variables for determining the PA's position, thereby introducing epistemic tension into the discussion. For example, if both human players choose the same stance, the PA adopts the opposing position to introduce tension in the discussion. Conversely, if human players disagree with each other, the PA aligns with the player who shows a lower opinion strength in their stance, amplifying the minority viewpoints. This adaptive positioning mechanism is designed to create meaningful socio-cognitive conflict and foster deeper engagement.

\subsection{Peer Agent Architecture}
Our PA has been designed with two main goals: First, the PA seamlessly integrates into game play as a socially coherent player and is perceived as a meaningful peer by human players. Second, the PA should dynamically adapt its intervention strategy to facilitate cognitive conflict during gameplay. This involves adopting an opposing stance and nudging human players to elaborate and defend their positions. Our architecture was informed by the Inner Thoughts framework \cite{liu2025proactive}, which enables proactive agent interventions through a turn-taking mechanism grounded in intrinsic motivation. Additionally, we extend this framework with a context-sensitive scoring system that guides the PA's behaviors in response to player dynamics. The proposed PA architecture is composed of five core modules: \textit{(1) Context Interpreter, (2) Agent State Manager, (3) Thought Generator, (4) Thought Evaluator, and (5) Thought Articulator.}
\paragraph{Context Interpreter} To participate in deliberation as a peer, the PA must recognize the context of both the game play and the ongoing discussion. The \textit{Context Interpreter} module stores each human player's position and confidence score for the given dilemma, forming the basis of their 'opinion state'. Additionally, players' dialogues are transcribed and analyzed based on the theory of basic human values \cite{schwartz2012overview} to detect value-laden content in the discussion. The module also monitors the current phase of the discussion (e.g., early vs. late rounds) to inform the selection and prioritization of discourse strategies.
 \paragraph{Agent State Manager}  Several agent states and traits are managed throughout gameplay: \textit{Position, Opinion Strength, Long-Term Memory, and Agent Persona}. Among these, opinion strength is a core state of the agent and determines whether human players' arguments persuade the PA. An agent's initial opinion strength is based on the average of the two human players' confidence scores. Based on prior work that LLMs can detect humans' persuasive arguments and update their inner opinion state \cite{bozdag2025persuade}, the agent's opinion strength is dynamically adjusted when persuasive arguments are detected in player conversations. This module also stores long-term memories and agent persona attributes, which ensure coherent behaviors and consistent thought generation throughout the discussion.
 \paragraph{Thought Generator}  Based on the player's current utterance, extracted values, and retrieved memory, this module generates multiple candidate thoughts that are categorized into two types \textit{General Thoughts and Strategic Thoughts}. \textit{General thoughts} refer to domain-general inner thoughts that humans use in daily conversations. \textit{Strategic thoughts} refer to thoughts intentionally constructed to deepen deliberations, based on prior work on transactive talk moves \cite{nguyen2023role, de2025investigating}. By distinguishing between these categories of thoughts, the agent can balance natural self-expression with deeper deliberation through intentional talk moves.
 \paragraph{Thought Evaluator} Once candidate thoughts are generated, the Thought Evaluator assesses each one by rating a motivation score on a 1-5 scale, reflecting the PA's intrinsic motivation to express the thought. This process is informed by an LLM-based reasoning framework and follows Liu et al's intrinsic motivation scoring model \cite{liu2025proactive}. The score is operationalized through a heuristic-based scoring mechanism, such as relevance, information gap, and expected impact. To further consider the appropriateness of strategic thoughts, rule-based logic is applied based on the discussion phases \cite{vogel2023transactivity} and the estimated opinion strength of human players to ensure that interventions are both timely and contextually sensitive.
  \paragraph{Thought Articulator} The Thought Articulator converts inner thoughts into a natural utterance with a peer-like tone. If a candidate thought's motivation score surpasses a pre-defined threshold, the strategic thought with the highest motivation score is selected. Otherwise, a general thought is probabilistically selected to maintain the conversational fluidity. The agent's proactivity and peer-likeness can be fine-tuned by adjusting the motivation threshold and the probability of selecting general thoughts.

\section{ONGOING WORK}
To gather early feedback, we developed a single-party prototype of the PA. A preliminary user study was conducted with four participants (1 male, 3 female) with computer science backgrounds. After interacting with the PA, participants evaluated it using a 7-point scale on two constructs of the social agency framework: perceived homophily and competence \cite{gong2008social}. The results indicated that participants perceived the PA's competence comparable to their own (Competence: M = 4.00, SD = 1.37), while recognizing that the agent expressed different viewpoints (Perceived Homophily: M = 3.38, SD = 1.73). Post-hoc interview yielded some key insights for future development: \paragraph{Need for Nuanced Expression} One participant (P2) noted that the agent's sudden shift in opinion disrupted the conversational flow. This highlights the need for the agent to more explicitly acknowledge opinion shifts, especially influenced by persuasive user arguments. Clarifying such shifts may maintain conversational coherence and social believability.
\paragraph{Need for an External Knowledge Base} Participants (P1 and P3) observed that some responses lacked sufficient supporting grounds, likely due to the initial prototype's reliance on user-provided inputs. This suggests that integrating an external knowledge base could enhance the agent's reasoning with more contextually grounded support. In future work, we plan to develop a full multi-party version of the PA system with an integrated voice interface and explore how it facilitates deliberation as a thoughtful opponent by inducing cognitive conflict to promote reflective thinking in dilemma-based learning environments.

\begin{acks}
This work was supported by the National Research Foundation of Korea (NRF) grant funded by the Korea government (MSIT) (RS-2024-00350045).
\end{acks}

\bibliographystyle{ACM-Reference-Format}
\bibliography{references}

\end{document}